\documentclass[prl,english,twocolumn,amsmath,showpacs]{revtex4}
\usepackage[latin1]{inputenc}
\usepackage{graphicx}% Include figure files
\usepackage{color}
\usepackage{babel}

\begin{document}

\title{Point-contact spectroscopy of the nickel borocarbide superconductor YNi$_{2}$B$_{2}$C
in the normal and superconducting state}

\author{D. L.\ Bashlakov, Yu. G.\ Naidyuk, I.  K.\ Yanson}

\affiliation{B.\ Verkin Institute for Low Temperature Physics and Engineering,
National Academy of Sciences of Ukraine, 47 Lenin Ave., 61103, Kharkiv,
Ukraine}

\author{G.\ Behr,  S.-L. Drechsler, G.\ Fuchs, L. Schultz and D.\ Souptel }

\affiliation{Leibniz-Institut f\"{u}r Festk\"{o}rper- und
Werkstoffforschung Dresden e.V., Postfach 270116, D-01171 Dresden,
Germany}

\date{\today{}}

\begin{abstract}
Point-contact (PC) spectroscopy measurements of YNi$_{2}$B$_{2}$C
single crystals in the normal and superconducting (SC) state
(T$_{c}\simeq$15.4\,K) for the main crystallographic directions
are reported. The PC study reveals the electron-phonon interaction
(EPI) spectral function with dominant phonon maximum around
12\,meV and further weak structures (hump or kink) at higher
energy at about 50\,meV. No "soft" modes below 12\,meV are
resolved in the normal state. The PC EPI spectra are qualitatively
similar for the different directions. Contrary, directional study
of the SC gap results in $\Delta_{\rm{[100]}}\approx$ 1.5\,meV for
the a direction and $\Delta_{\rm{[001]}}\approx$ 2.3\,meV along
the c axis; however the critical temperature $T_{\rm c}$ in PC in
all cases is near to that in the bulk sample. The value
2$\Delta_{\rm [001]}$/$k_{\rm B}T_{\rm c}\approx$ 3.6 is close to
the BCS value of 3.52, and the temperature dependence $\Delta_{\rm
[001]}(T)$ is BCS-like, while the for small gap $\Delta_{\rm
[100]}(T)$ is below BCS behavior at $T>T_c/2$ similarly as in the
two-gap superconductor MgB$_2$. It is supposed that the
directional variation $\Delta$ can be attributed to a multiband
nature of the SC state in YNi$_{2}$B$_{2}$C.


\end{abstract}
\maketitle

\section{Introduction}

The family of borocarbide superconductors $RT_{2}$B$_{2}$C, where
$R$ is a rare-earth element and $T$ is a transition metal element
(mainly Ni) have been studied intensively after superconductivity in
$RT_{2}$B$_{2}$C was discovered in 1994 \cite{Cava,Nagarajan}.
Nevertheless, the nature and mechanism of superconductivity in
borocarbides which have 3D electronic structure are still under
debate. The most thermodynamic, transport and spectroscopic
measurements \cite{Muller,Muller1} give evidence that the
superconducting (SC) state has an $s$-wave symmetry and the pairing
is mediated by the electron-phonon interaction. However, for
nonmagnetic $R$ = Lu and Y compounds there are several properties
mentioned in Ref.\cite{Drechsler}, which when taken together, might
be interpreted also as hints for unconventional $d$-wave or $p$-wave
superconductivity. An anisotropic s-wave order parameter (or s+g
model) was proposed for LuNi$_{2}$B$_{2}$C and YNi$_{2}$B$_{2}$C in
\cite{Maki}. Anyhow, irrespective of the actual order parameter
symmetry, there is clear evidence for a notable anisotropy of the SC
gap in YNi$_{2}$B$_{2}$C
\cite{Izawa,Park,Martinez,Pratap,Mukho,Bashlakov,Yokoya,Huang} at
least on parts of the complex Fermi surface
\cite{drechsler04,drechsler03,drechsler01}.

Using point-contact (PC) spectroscopy \cite{Naid} both the SC
order parameter and the PC electron-phonon interaction (EPI)
function $\alpha_{\textrm{PC}}^{2}F(\omega)$ can be determined
from the first and second derivatives of the $I(V)$ characteristic
of PC, respectively. The measurement of the second derivative of
the $I(V)$ for PC provides straightforward information as to the
PC EPI function $\alpha_{\textrm{PC}}^{2}F(\omega)$
\cite{Naid,KOS,Kulik}. The knowledge of $\alpha^{2}F(\omega)$ for
conducting systems is a touchstone as to phonon-mediated
superconductivity, which is governed by the value of the
electron-phonon-coupling parameter
$\lambda=2\int\alpha^{2}F(\omega)\omega^{-1}d\omega$. Moreover,
the comparison of the experimentally determined
$\alpha^{2}F(\omega)$ with the calculated one can discriminate
different theoretical models and approaches. Thus the PC
spectroscopy could be helpful to illuminate details of the EPI in
$RT_{2}$B$_{2}$C as well as to resolve non-phonon quasiparticle
interactions. In addition, the SC gap determines the behavior of
the $I(V)$ curve of PC at the low biases of a few mV, which is
widely used to derive the SC gap from routine fitting by the
well-known Blonder-Tinkham-Klapwijk (BTK) equations \cite{BTK82}.
These experimental information as to the SC gap and EPI function
is very useful for understanding the SC properties and the
mechanism of superconductivity for the material under study.

Up to now there are a few papers where YNi$_{2}$B$_{2}$C has been
investigated by PC spectroscopy
\cite{Ryba1,Pratap,Mukho,Bashlakov,Yanson0,Yanson1}. The first
four papers are devoted to study of the SC gap, while the PC EPI
spectra for YNi$_{2}$B$_{2}$C were measured and analyzed in
\cite{Yanson0,Yanson1}. The main attention there was devoted to
study the low energy part of the PC spectra and to so called
"soft" mode at about 4-5\,meV in the electron-quasiparticle
spectrum. However, the measured PC spectra were featureless above
20\,mV, although a number of pronounced phonon peaks are well
resolved at higher energy by neutron spectroscopy \cite{Gompf}. In
this paper we present more detailed data as to PC EPI spectra of
YNi$_{2}$B$_{2}$C and as to directional measurements of the SC gap
in this compound.  The data were reported on M$^2$S-HTSC
Conference in Dresden (July 9-14, 2006) and partially published in
Physica C \cite{M2Sconf}.

\section{Experimental details}

We have used single crystals of YNi$_{2}$B$_{2}$C grown by a
floating zone technique with optical heating \cite{Souptel}. The
sample has a residual resistivity of
$\rho_{0}\simeq1\mu\Omega$\,cm and a residual resistivity ratio
(RRR) about 40. It  becomes superconducting at about 15.4\,K with
a transition width about 0.1\,K. PCs were established both along
the c axis and along the a axis as well as in the basal plane
close to the [110] direction by standard "needle-anvil" or "shear"
methods \cite{Naid}. As a counter electrode Cu or Ag thin
($\oslash\simeq$0.15\,mm) wires were used to study the SC gap via
the mechanism of Andreev reflection. A series of measurements were
done using homocontacts between two pieces of YNi$_{2}$B$_{2}$C.
In this case the orientation of the contact axis with respect to
the crystallographic directions was not controlled. The
experimental cell was placed in a flow cryostat, enabling
measurements from 1.5\,K up to $T_{\rm c}$ and higher. PCs were
created by touching the YNi$_{2}$B$_{2}$C surface by sharpened Cu
and Ag wires directly in the cryostat at liquid helium
temperature. To establish homocontacts two pieces of
YNi$_{2}$B$_{2}$C were touched. An disadvantage of the
"needle--anvil" method is the sensitivity of the contacts to
mechanical vibrations and to change of temperature. As a result,
temperature measurements have sustained about a quarter from total
in four tens of the contacts investigated at liquid helium
temperature.

Using a technique of synchronous detection of weak alternating
signal harmonics, the first harmonic of the modulating signal $V_1$
(proportional to the differential resistance $R(V)=dV/dI(V)$) and
the second harmonic $V_2$ (proportional to $d^2V/dI^2(V)$) were
recorded as a function of the bias voltage $V$. $V_2(V)$ can be
expressed as follows:
\begin{equation}
\label{v2} V_2(V)= \frac{V_1^2}{2\sqrt{2}}R_0^{-1}\frac{dR(V)}{dV}.
\end{equation}
According to the theory of PC spectroscopy \cite{KOS,Kulik} the
second derivative $R^{-1}dR/dV=R^{-2}d^{2}V/dI^{2}(V)$ of the
$I(V)$ curve of the ballistic contact at low temperatures is
determined by the PC EPI function
$\alpha_{\textrm{PC}}^{2}F(\epsilon)$:
\begin{equation}
\label{pcs} R^{\rm -1}\frac{{\rm d}R}{{\rm d}V}= \frac{8\,{\rm
e}d}{3\,\hbar v_{\rm F}}\alpha_{\rm
PC}^2(\epsilon)\,F(\epsilon)|_{\epsilon={\rm e}V} ,
\end{equation}
where e is the electron charge,  $d$ is the PC diameter and
$\alpha_{\textrm{PC}}$, roughly speaking, reflects strength of the
interaction of electrons with phonons. This interaction underlines
the large-angle scattering (back-scattering) processes \cite{KOS} of
electrons in the PC constriction. Thus
$\alpha_{\textrm{PC}}^{2}F(\epsilon)$ is a kind of transport EPI
function which selects phonons with a large momentum or Umklapp
scattering. The PC diameter $d$, which enters in Eq.(\ref{pcs}), can
be calculated by the Wexler \cite{Wexler} relation:
\begin{equation}
R_{\rm PC}(T) \simeq  \frac {16 \rho l}{3\pi d^2} + \frac{\rho
(T)}{d}, \label{wex}
\end{equation}
which consists of a sum of ballistic (Sharvin) and the diffusive
(Maxwell) terms. Here $\rho l = p_{\rm F}/n$e$^2$ corresponds to the
free electron model, where $p_{\rm F}$ is the Fermi momentum and $n$
is the density of charge carriers. Using data for
$\rho=2.5\mu\Omega$\,cm and $l$=41\,nm from \cite{Shulga} the
product $\rho l$ is about 10$^{-11}\Omega$\,cm$^2$ in
YNi$_{2}$B$_{2}$C.

From (\ref{v2}) and (\ref{pcs})
$\alpha_{\textrm{PC}}^{2}(\epsilon)\, F(\epsilon)$ can be defined
as:
\begin{equation}
\label{pcs1} \alpha_{PC}^{2}(\epsilon)\,F(\epsilon)=
\frac{3}{2\sqrt{2}}\frac{\hbar v_{\rm F}}{{\rm
e}d}\frac{V_{2}}{V_{1}^{2}},
\end{equation}
that is the measured ac voltage $V_2$ weighted by $V_1^2$ is
directly proportional to the PC EPI function
$\alpha_{\textrm{PC}}^{2}(\epsilon)\, F(\epsilon)$. Here it is
necessary to mention that the finite temperature $T$ and the
alternating voltage $V_1$  result in a smearing of the measured
$V_2$ spectra. Thus, the infinitely narrow spectral peak smears
into a bell-shaped maximum (see, e.\,g. \cite{Naid}) with the
width
\begin{equation}
\delta = [(5.44\,{\rm k_B}T/{\rm e})^2 + (1.22
\sqrt2\,V_1)^2]^{1/2}. \label{kT}
\end{equation}
For example, the smearing of a PC spectrum measured at liquid
helium temperature 4.2\,K and at ac voltage $V_1$ between 1 and
2\,mV, mainly used in our measurements, is between 2.6 and 4\,mV.

The PC EPI function $\alpha_{PC}^{2}(\epsilon)\,F(\epsilon)$
should vanish above the maximum phonon energy $\hbar\omega_{\rm
max}$, which is close to the Debye energy $k_{\rm B}T_{\rm D}$,
because of the lack of phonons with larger energy. Therefore,
according to (\ref{pcs1}) the PC spectrum $V_2(V)$ should vanish
above $eV\ge \hbar\omega_{\rm max}$. In fact,  measured PC spectra
always have a nonzero almost constant value above the Debye
energy, the so-called background. The general nature of the
background was understood by taking into account an accumulation
of nonequilibrium phonons in the PC region created by the
energized electrons. The details of the background calculations
are given elsewhere \cite{Kulik,Kulik1}. The most often
semiempirical formula \cite{Naid}
\begin{equation}
\label{bg} B(eV) ={\rm const}\, \int_0^{eV}\frac{\alpha_{\rm
PC}^2(\epsilon)F(\epsilon)}{\epsilon}\rm{d}\epsilon
\end{equation}
is used to describe the background. Here for calculations of the
energy dependent background an iterative procedure is applied taking
as a first approximation for $B(eV)$  a curve, which continuously
increases from zero to the maximal background value at $eV\ge
\hbar\omega_{\rm max}$.

In the case of a heterocontact between two metals the PC spectrum
represents a sum of the contributions from both metals 1 and 2
weighted by the inverse Fermi velocity \cite{Shekhter83}:
\begin{equation}
\frac{V_{2}}{V_{1}^{2}}\propto
\upsilon\frac{(\alpha^{2}F)_{1}}{v_{F1}}+(1-\upsilon)\frac{(\alpha^{2}F)_{2}}{v_{F2}},
\label{het}
\end{equation}
where $\upsilon$ is the relative volume occupied by metal 1 in the
PC. Thus, using of  heterocontacts enables also, e.\,g.,
qualitative estimation of the relative strength of EPI in the
investigated material as compared to some standard or well known
one.

According to the BTK theory \cite{BTK82} of conductivity of N-c-S
heterocontacts (here N is the normal metal, c is the constriction
and S is the superconductor) a maximum at zero-bias voltage and a
double-minimum structure at about $V\simeq \pm\Delta /$e in the
d$V/$d$I$ curves manifest the Andreev reflection processes at the
N-S interface with a finite, so called,  barrier strength
parameter $Z\neq$0. The latter has a simple interpretation: it
increases normal state resistance of the PC by a factor of
(1+$Z^2$). Thus, as it was mentioned, the minima position in
d$V/$d$I$ reflect roughly the SC gap value, which follows from the
equations for the $I(V)$ characteristics \cite{BTK82}:
\begin{eqnarray}\label{BTKeq}
I(V) &\sim &\int_{-\infty }^{\infty}
T(\epsilon)\left(f(\epsilon-{\rm e}V)-f(\epsilon)\right) {\rm
d}\epsilon,  \\ T(\epsilon) &=&\frac{2\Delta ^2}{\epsilon^2+(\Delta
^2-\epsilon^2)(2Z^2+1)^2},~~~|\epsilon|<\Delta \nonumber \\
T(\epsilon) &=&\frac{2|\epsilon|}{|\epsilon|+\sqrt{\epsilon^2-\Delta
^2}(2Z^2+1)},~~~ |\epsilon|>\Delta ~, \nonumber
\end{eqnarray}
where f($\epsilon$) is the Fermi distribution function. In
general, $Z$ characterizes reflection (or transmission) of the N-S
interface, which is defined also by the mismatch of the Fermi
velocity $v_{F}$. Thus, even in the absence of a "natural" barrier
$Z$ is non zero and is given by
\begin{equation}
Z=\frac{|v_{F_1}-v_{F_2}|}{2(v_{F_1}v_{F_2})^{1/2}}~~.  \label{Z}
 \end{equation}
The smearing of the experimental $dV/dI$ curves as compared to the
calculated ones according to (\ref{BTKeq}) is usually attributed to
quasiparticle DOS N($\epsilon,\Gamma $) broadening in the
superconductor due to finite-lifetime. According to Dynes et al.
\cite{Dynes78} it can be taken into account by adding an imaginary
part to the energy $\epsilon$, namely,
%\begin{equation}
%N(\epsilon,\Gamma )={\rm Re}\left\{ \frac{\epsilon-i\Gamma }
%{\sqrt{(\epsilon-i\Gamma)^2-\Delta ^2}}\right\}, \label{Dyneseq}
%\end{equation}
$\epsilon$ is replaced by $\epsilon-i\Gamma$ in (\ref{BTKeq}). We
used (\ref{BTKeq}) to fit the measured d$V/$d$I$ curves of PCs and
to extract the SC gap.

\section{Experimental results and discussion}

\subsection{PC spectroscopy of quasiparticle excitations}

\begin{figure}
\begin{center}
\includegraphics[width=8cm,angle=0]{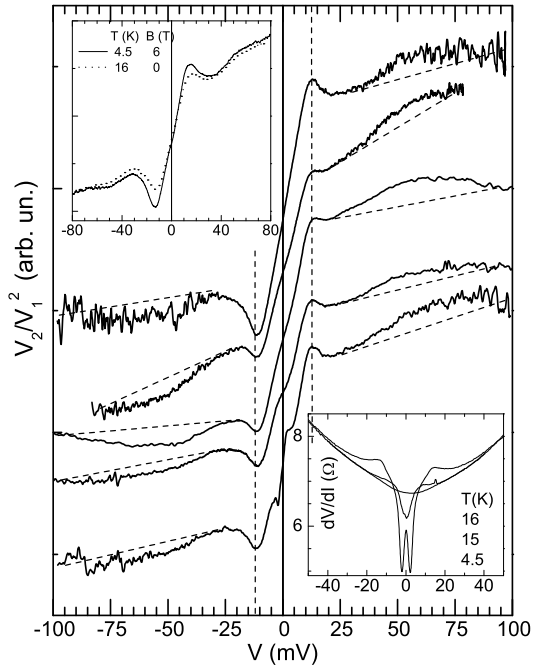}
%\includegraphics[
%  width=9.5cm,bb = 0 0 200 100, draft, type=eps]{y1.eps}
\end{center}
\caption{PC spectra {[}see Eq.\,(\ref{pcs1}){]} of several
YNi$_{2}$B$_{2}$C--Ag contacts with resistance 8.8, 6.8, 3.1, 2.8
and 5.1$\Omega$ (from the top curve to the bottom one). PC spectra
are measured at 16\,K $>$T$_c\simeq$15.4\,K to suppress
superconductivity and avoid huge features in $V_2$ between $\sim
\pm $20\,mV due to the gap minimum shown in the bottom inset. The
bottom curve (measured at 15.1\,K -- slightly below T$_c$)
demonstrates these almost suppressed (sharp kink/peak) features
close to zero bias. Vertical dashed lines are drawn to help to
follow position of the main maximum in the PC spectra and tilted
lines are drawn to accentuate hump around 50\,mV. Upper inset
shows PC spectra of YNi$_{2}$B$_{2}$C--Cu contacts with resistance
4.5\,$\Omega$ measured at $V_1(0)$=2\,mV in the normal state by
suppressing superconductivity via temperature or magnetic field.}
\label{y1}
\end{figure}
\begin{figure}
\begin{center}
\includegraphics[width=7cm,angle=0]{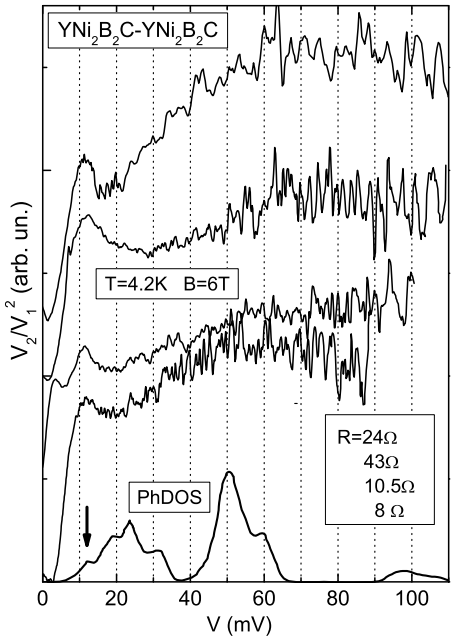}
\end{center}
\caption{PC spectra {[}see Eq.\,(\ref{pcs1}){]} of several
YNi$_{2}$B$_{2}$C--YNi$_{2}$B$_{2}$C homo contacts with different
resistance (shown in inset). The PC spectra are measured in magnetic
field 6\,T to suppress SC features at low bias. The PC spectrum for
10.5\,$\Omega$ contact still shows maximum around 4\,mV due to
residual superconductivity. The bottom curve shows phonon density of
states (PhDOS) for YNi$_{2}$B$_{2}$C from \cite{Gompf}.} \label{y2}
\end{figure}
\begin{figure}
\begin{center}
\includegraphics[width=7cm,angle=0]{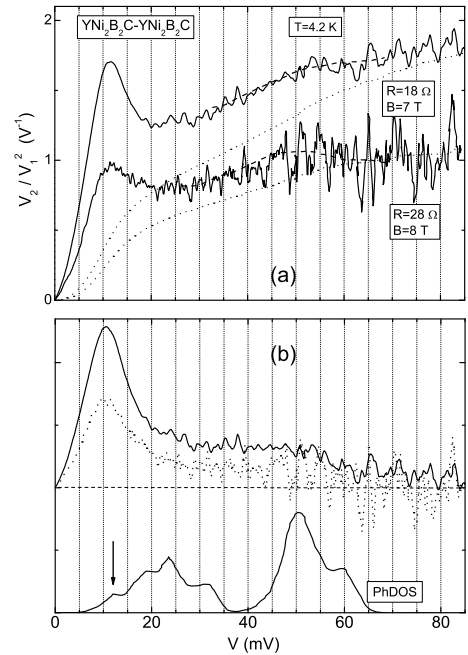}
\end{center}
\caption{(a) PC spectra {[}see Eq.\,(\ref{pcs1}){]} of two
YNi$_{2}$B$_{2}$C homocontacts averaged for two polarities. The
superconductivity is suppressed by a magnetic field. Dotted curves
show the background behavior calculated according to (\ref{bg}).
Dashed curves on the PC spectra are drawn by hand to better
visualize hump around 50\,mV. $V_1(0)$ is 0.7 and 0.87\,mV, which
gives $\delta$=2.3 and 2.5\,mV according to (\ref{kT}). (b) PC
spectra from the upper panel after subtracting of the background.
The bottom curve shows PhDOS for YNi$_{2}$B$_{2}$C \cite{Gompf}.
Vertical arrow shows position of the maximum in PhDOS which
corresponds to the main maximum in the PC spectra.} \label{y3}
\end{figure}

Figure \ref{y1} shows PC spectra of several YNi$_{2}$B$_{2}$C--Ag
heterocontacts for which the SC gap has been measured
simultaneously. We have selected the PC spectra for which the SC gap
varies (see Fig.\,\ref{y4}) gradually from the maximal value of
2.5\,meV (upper spectrum) to the minimal one of 1.65\,meV (bottom
spectrum). However, no qualitative difference between PC EPI spectra
is observed. The spectra show a dominant maximum at about 12\,mV and
a broad shallow maximum or hump centered around 50\,mV
(Fig.\,\ref{y1}). These maxima correspond well to the phonon DOS
maxima at 12 and 50\,mV of YNi$_{2}$B$_{2}$C (see Fig.\,\ref{y2})
obtained by neutron diffraction \cite{Gompf}. At the same time the
PC spectra do not contain contributions from the other phonon maxima
at 20, 24, 32\,mV and 100\,mV observed in the phonon DOS. In this
context we should mention that PC spectra of HoNi$_{2}$B$_{2}$C
\cite{NaidSCES} display mentioned phonon maxima near 20, 24, 32\,mV.
Contrary, the PC spectra of YNi$_{2}$B$_{2}$C above 12\,mV is
monotonic and almost featureless, except of the mentioned 50\,mV
feature. Of course, the increase of noise with the voltage hides
details of the spectra at higher energies. Note also, that the
spectra in Fig.\,\ref{y1} are measured in the normal state at 16\,K
where the resolution due to the Fermi level smearing is according to
(\ref{kT}) about 8\,meV which can mask fine features. However, as
the upper insert in Fig.\,\ref{y1} shows, the spectrum measured at
4.5\,K (solid curve) with the resolution of about 4\,meV is similar
to the other one (dashed curve) only it has a little bit sharper
maximum at 12\,mV. Even improving resolution below 3\,mV (see
Fig.\,\ref{y3}) does not recover additional details of the spectra.
It means that the instrumental broadening of the spectra does not
play here a crucial role. In this respect we should note that
according to recent data \cite{Reich} the strong EPI gives rise to
pronounced anomalies in the phonon dispersion curves of
YNi$_{2}$B$_{2}$C and concurrently to {\it large line widths} of
certain phonon modes. The latter along with selection of the
large-angle scattering processes in PC can be responsible for the
broad and less detailed (compared to PhDOS) structure in PC EPI
spectra.

By interpreting the PC spectra of heterocontacts we have to take
into account for possible contributions of the normal metal (e.g.,
Ag or Cu) used as a counter electrode. To avoid this we have
measured PC spectra  of homocontacts shown in Fig.\,\ref{y2}. No
qualitative difference in the spectra of homo and heterocontacts
is seen (compare spectra in Figs.\,\ref{y1} and \ref{y2}). Thus,
the contribution of Ag or Cu in the presented PC spectra of
heterocontacts (see, Fig.\,\ref{y1}) is negligible. Apparently,
low Fermi velocities in nickel borocarbides \cite{Muller,Muller1}
as compared to the noble metals accentuate the contribution of
YNi$_{2}$B$_{2}$C in the PC spectra according to Eq.\,(\ref{het}).

Fig.\,\ref{y3} displays the PC spectra of two homocontacts
averaged for negative and positive polarity. Here also the
behavior of the background obtained from (\ref{bg}) and the
neutron phonon DOS are shown. According to neutron data
\cite{Gompf} there is a gap around 40\,meV, which separates
acoustic and optic phonon branches. In this energy region only a
flattering occurs in our PC spectra after subtracting of the
background. The main reason is that subtraction of the background
for PC spectra with a high background level is not a
straightforward procedure. Nevertheless, the PC spectra show two
remarkable features -- a maximum at about 12\,mV and a hump around
50\,mV. Similar PC EPI spectra of YNi$_{2}$B$_{2}$C were presented
in \cite{Yanson0,Yanson1}. Here we only note that the measured
{\it normal} state PC EPI spectra of YNi$_{2}$B$_{2}$C demonstrate
no "soft" modes around 4--5\,mV discussed by Yanson et al.
\cite{Yanson0,Yanson1} and noticed by Martinez-Samper et al.
\cite{Martinez} in STM spectra. Similar to "soft" mode maxima are
seen in the bottom spectrum in Fig.\,\ref{y1} and in the spectrum
of 10.5\,$\Omega$-contact in Fig.\,\ref{y2}, but they can be
attributed to superconductivity which is not fully suppressed in
these contacts.

After subtracting the background from the measured PC spectra the
EPI function is established according to Eq.(\ref{pcs1}) (see
Fig.\,\ref{y3}b) and the EPI parameter
$\lambda=2\int\alpha^{2}F(\omega)\omega^{-1}d\omega$ is
calculated, which is found to be about 0.1. However, the
calculation of $\lambda$ from a PC spectrum is complicated for
several reasons. First of all equation (\ref{pcs1}) is derived for
a free electron model and a single band Fermi surface. Secondly,
deviations from the ballistic regime in PC due to elastic
scattering have to be corrected by a pre-factor $l_{i}/d$ in
(\ref{pcs1}), where $l_{i}$ is the elastic electron mean free path
and $d$ is the PC diameter. However, $l_{i}$ is difficult to
evaluate for the PC. In this case only a qualitative estimation of
$\lambda$ is possible taking into account that contribution of Ag
or Cu in the PC spectra of YNi$_{2}$B$_{2}$C--Ag/Cu heterocontacts
in Fig.\,\ref{y1} (PC spectra of YNi$_{2}$B$_{2}$C--Cu
heterocontact is shown in inset) is hardly to resolve. From the
latter we can conclude that the intensity of the EPI function in
YNi$_{2}$B$_{2}$C is at least larger than that in Cu, where
$\lambda\simeq$0.25 \cite{Naid}. This provides an complementary
confirmation of the moderate EPI in YNi$_{2}$B$_{2}$C with the
lower limit of 0.25 for the $\lambda$ \footnote{By discussion of
the $\lambda$ value calculated from PC spectra we should always to
bear in mind that $\lambda_{PC}$ is some kind of transport
$\lambda$ and it can be, in general, different from the
thermodynamic one. Moreover, as we have discussed in the case of
HoNi$_{2}$B$_{2}$C compound (Yu. G. Naidyuk et al., will be
published) different bands can have different $\lambda$, while
from PC spectra some average $\lambda_{PC}$ is calculated.}. Note
also that because the 12\,mV-maximum prevails in the PC spectra,
the main (about 90\%) contribution to $\lambda_{PC}$ comes from
the energy region below 35\,meV corresponding to the low energy
(acoustic) part of the phonon DOS. Calculation of EPI coupling in
YNi$_{2}$B$_{2}$C revealed that about 70\% of $\lambda$ results
from the nine lowest branches \cite{Reich}.

No significant anisotropy of the PC EPI spectra in YNi$_{2}$B$_{2}$C
was observed (see, e.g., Fig.\,\ref{y1}). The main reason can be
that the spectra are quite broad and smeared, what hides the fine
structure of EPI, which might be anisotropic. However, isotropic
behavior of the resistivity in YNi$_{2}$B$_{2}$C mentioned in
\cite{Muller,Muller1} is in line with almost isotropic PC spectra.

\subsection{Directional PCS of the SC energy gap}

\begin{figure}
\begin{center}
\includegraphics[width=8cm,angle=0]{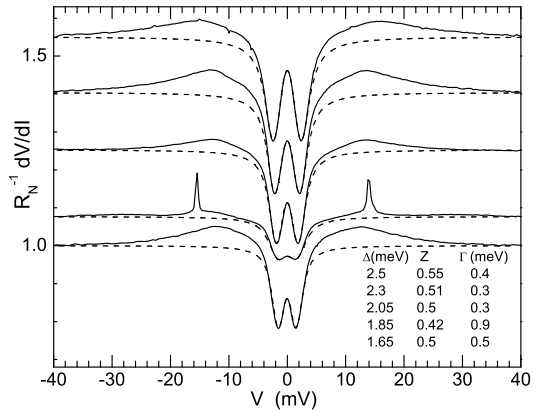}
\end{center}
\caption{Reduced $dV/dI$ curves (solid lines) of
YNi$_{2}$B$_{2}$C--Cu contacts at $T$=4.2\,K. PC spectra for these
contacts are shown in Fig.\,\ref{y1}. Dashed lines are BTK fitting
curves according to (\ref{BTKeq}). The curves are shifted
vertically versus the bottom one for clarity. The table shows SC
energy gap $\Delta$, $\Gamma$ and $Z$ parameters obtained by BTK
fitting of the experimental curves. } \label{y4}
\end{figure}

As it was mentioned above the SC gap manifests itself in the
$dV/dI$ characteristic of a N-c-S contact as minima around
$V\simeq\pm\Delta$ if $Z\neq$0 and a temperature is well below
T$_{c}$. Such $dV/dI$ curves are presented in Fig.\,\ref{y4} for
several contacts whose PC spectra are shown in Fig.\,\ref{y1}.
$dV/dI$ in Fig.\,\ref{y4} reflects also the distribution of the
gap in YNi$_{2}$B$_{2}$C.

\begin{figure}
\begin{center}
\includegraphics[width=7cm,angle=0]{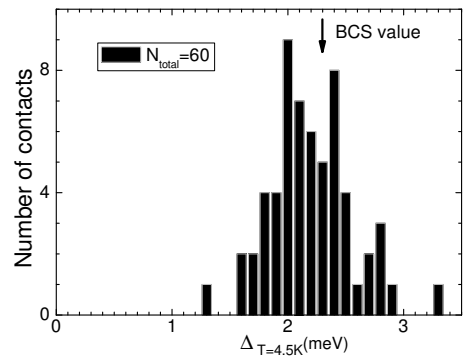}
\end{center}
\vspace{-1cm} \caption{Gap distribution measured for a cleaved
nonoriented surface of YNi$_{2}$B$_{2}$C single crystal. }
\label{y5a}
\end{figure}
\begin{table}
\caption[]{Average, minimal and maximal values of the SC gap
$\Delta$, `smearing` parameter $\Gamma$ and  `barrier` parameter $Z$
for PCs represented gap distribution in Fig.\,\ref{y5a}.}
\begin{tabular}{|c|c|c|c|}
  \hline
  % after \\: \hline or \cline{col1-col2} \cline{col3-col4} ...
   & $\Delta$(meV) & $\Gamma$(meV) & Z \\
  \hline
\textbf{Average} & \textbf{ 2.2} & \textbf{ 0.64} & \textbf{ 0.5} \\
    Minimal & 1.3 & 0.3 & 0.33 \\
      Maximal & 3.3 & 2.3 & 0.69 \\
  \hline
\end{tabular}
\end{table}
\begin{figure}
\begin{center}
\includegraphics[width=8cm,angle=0]{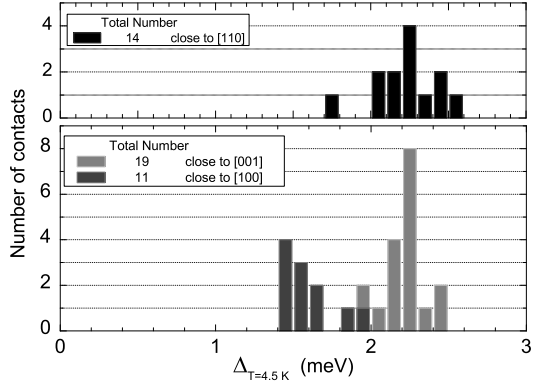}
\end{center}
\caption{Gap distribution for the three main directions in
YNi$_{2}$B$_{2}$C single crystal. In comparison with our gap
distribution presented in \cite{M2Sconf}, here, the data are added
measured for 21 contact for YNi$_{2}$B$_{2}$C with improved RRR
$\simeq$ 60. But it makes no qualitative changes, only improves
statistic. } \label{y5b}
\end{figure}

The gap distribution is also shown in Fig.\,\ref{y5a} for a
cleaved nonoriented rough YNi$_{2}$B$_{2}$C surface. The
characteristic values of the fitting parameters are shown in Table
I. The distribution in Fig.\,\ref{y5a} is similar to that observed
for YNi$_{2}$B$_{2}$C films \cite{Bashlakov}. Different from the
films is that the gap values in Fig.\,\ref{y5a} stretched above
2.4\,meV. Note, that the average gap is close to the BCS value of
$\Delta$=1.76$k_BT_c\simeq$2.3\,meV. Interesting that using
effective Fermi velocities in YNi$_{2}$B$_{2}$C in the range from
0.45$\times10^5$ to 4.5$\times10^5$\,m/s \cite{drechsler01} within
the two-band model \cite{Shulga} and a typical Fermi velocity of
Cu 1.57$\times10^6$\,m/s, the barrier parameter $Z$ can be
estimated according to (\ref{Z}) to be between 0.67 and 2.9
values, that is the lower value is close to the maximal Z from the
Table I.

The gap distribution for different crystallographic orientations in
YNi$_{2}$B$_{2}$C is shown in Fig.\,\ref{y5b}. The anisotropy in the
distribution is clearly seen: a small gap is characteristic for the
a-axis, while along the c-axis the gap is larger. Also the [110]
direction has in average a slightly enhanced  gap.

\begin{figure}
\begin{center}
\includegraphics[width=8cm,angle=0]{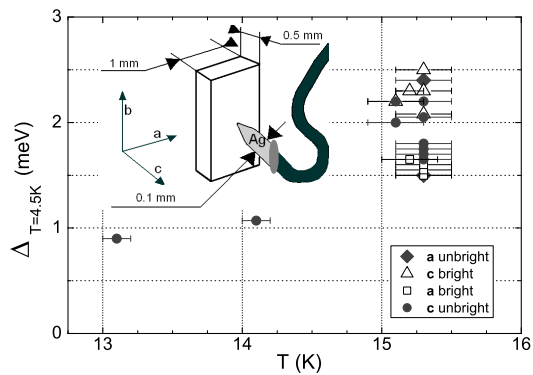}
\end{center}
\vspace{-0.5cm} \caption{Gap distribution for the polished
(bright) and unpolished (unbright) surface of YNi$_{2}$B$_{2}$C
single crystal for two main directions versus critical temperature
in PC. Inset shows draft of the YNi$_{2}$B$_{2}$C single crystal
and sharpened Ag wire.} \label{y6}
\end{figure}

To exclude the gap variation due to surface degradation we have
checked also the critical temperature of the gap vanishing for
most of PCs. In Fig.\,\ref{y6} gap values for the a and c axis for
the surface before polishing (marked by dark symbols) and after
mechanical polishing and chemical etching (marked by bright
symbols) are plotted against the critical temperature T$_c$ of the
PC. It is seen that for two PCs with low gap value of about 1\,meV
T$_c$  is reduced as compared to the bulk one, therefore, a
degradation of the SC state at the surface is likely responsible
for the low gap value ($\leq$ 1\,meV) rather than an anisotropy of
the gap. Contrary, different gap values for PCs with the same
T$_c$ shown in Fig.\,\ref{y6} gives unequivocal evidence of
intrinsic reason of the gap variation.

It is noted that the derived gap values are fairly consistent with
recent specific heat data \cite{Huang}. Here it was shown that the
two-gap model with $\Delta$=2.67\,meV and 1.19\,meV describe the
SC gap function of YNi$_{2}$B$_{2}$C better than other models
based on the isotropic s-wave, the d-wave line nodes, or the s+g
wave approach. Furthermore, as it was shown in \cite{Bobrov}
two-gap fit better describes d$V$/d$I(V)$ curve of PCs in the
sister compound LuNi$_{2}$B$_{2}$C.

\begin{figure}
\begin{center}
\includegraphics[width=8cm,angle=0]{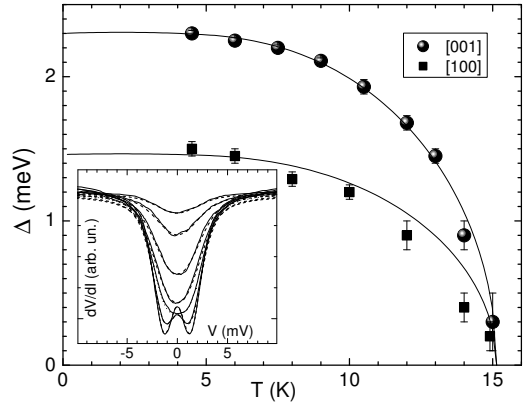}
\end{center}
\caption{Temperature dependence of the small (1.5\,meV) and large
(2.3\,meV) gap in YNi$_{2}$B$_{2}$C. Solid curves represent BCS-like
behavior. Inset shows normalized to the normal state $dV/dI$ curves
(solid) for the small gap along with calculated curves (dashed)
according to (\ref{BTKeq}). } \label{y7}
\end{figure}

The temperature dependence of $\Delta$ for two PCs with different
gaps is shown in Fig.\,\ref{y7}. It is seen that $\Delta(T)$ has in
general  BCS-type dependence. The gap is found to vanish close to
the bulk T$_{c}$. However, the small gap deviates from the BCS curve
by approaching T$_{c}$. Similar (small) gap behavior is
characteristic for the well-known multiband (two-gap) superconductor
MgB$_2$ \cite{Szabo}. Thus the mentioned observations of the gap
behavior and distribution can be taken as support of two-gap
scenario in YNi$_{2}$B$_{2}$C. At the same time as follows from
recent ARPES experiments \cite{Yokoya}, the momentum-dependent
superconducting gap shows a large anisotropy ($\Delta=2.3-3.2$\,meV)
observed on a single FS. Also ultrahigh-resolution photoemission
spectra \cite{Baba} are better described by anisotropic s-wave gap
in the form $\Delta(\theta)=2.8|\cos\, 2\theta|$\,(meV). Therefore,
the gap behavior in YNi$_{2}$B$_{2}$C is rather complex, showing
both anisotropic and multiband superconductivity.

\subsection{Magnetic field behavior of the SC energy gap and excess current}

\begin{figure}
\begin{center}
\vspace{3cm}
\includegraphics[width=8cm,angle=0]{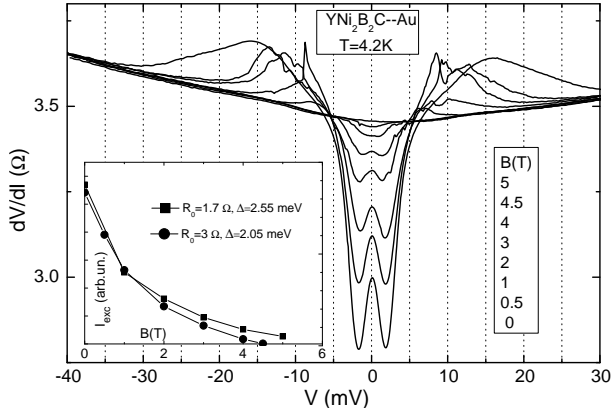}
\end{center}
\vspace{-2.5cm}
 \caption{Magnetic field dependence of $dV/dI$
curves for YNi$_{2}$B$_{2}$C--Au contact. Inset shows behavior of
the excess current for this contact and for another one with
larger $\Delta$. } \label{y8}
\end{figure}

\begin{figure}
\begin{center}
\includegraphics[width=8cm,angle=0]{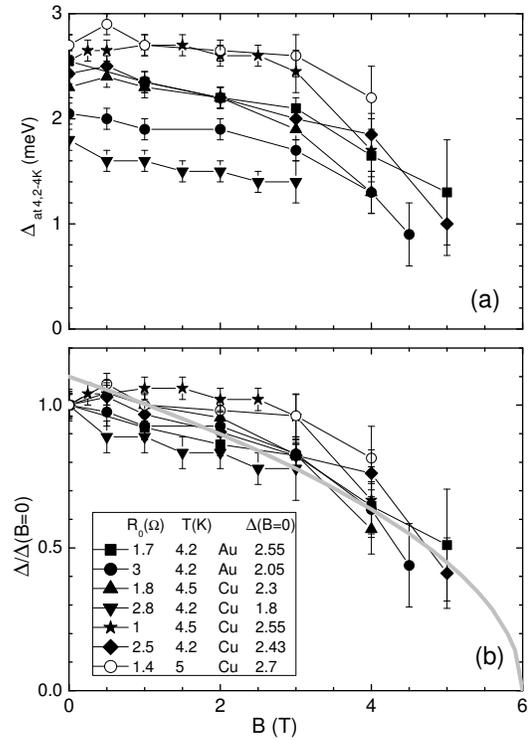}
\end{center}
\caption{Magnetic field dependence of the SC gap at 4.2\,K in
YNi$_{2}$B$_{2}$C extracted from $dV/dI$ for a few contacts: (a) SC
gap in absolute value, (b) SC gap reduced to $\Delta(B=0)$. Grey
curve shows the behavior $\propto(1-B/B_{c2})^{1/2}$ of the pair
potential of a type II superconductor in the vortex state according
to Abrikosov's theory.} \label{y9}
\end{figure}

In Fig.\,\ref{y8} the field dependence of $dV/dI$ curves for a
YNi$_{2}$B$_{2}$C--Au contact is shown. A remarkable suppression
of the minimum in $dV/dI$ is observed for applied field above
3\,T. This increases of the error in the determination of the
$\Delta$ value and makes calculations of $\Delta$ less accurate,
especially close to the critical field. The latter can be
estimated of about 6\,T from the SC features (minimum)
disappearing in $dV/dI$, that is it is close to the bulk critical
field at helium temperature 4.5\,K (see e.\,g. \cite{Shulga}).
$\Delta(B)$ behavior extracted from the $dV/dI$ curves for this
and several other contacts is shown in Fig.\,\ref{y9}. In general,
$\Delta(B)$ exhibits a conventional behavior decreasing with
overall negative curvature. At the same time the excess current
$I_{\textrm{exc}}$, which is roughly speaking proportional to the
area of the gap minimum in $dV/dI$ (or more precisely to the area
of the gap maximum in differential conductivity $dI/dV$),
decreases with the magnetic field with a positive curvature (see
Fig.\,\ref{y8} inset). Similar $I_{\textrm{exc}}(B)$ behavior we
have recently reported for PC on YNi$_{2}$B$_{2}$C film
\cite{Bashlakov} and before a remarkable positive curvature in
$I_{\textrm{exc}}(B)$ has been observed for the two-band
superconductor MgB$_{2}$ \cite{NaidyukF}. As it was shown for the
first time in \cite{Artemenko}, the excess current of an S-c-N
contact is governed by $\Delta$ or SC order parameter
\cite{Belob}. Indeed, the temperature dependence of
$I_{\textrm{exc}}(T)$ (not shown) is similar to the $\Delta(T)$
dependence, that is $I_{\textrm{exc}}(T)$ has a negative
curvature, while $I_{\textrm{exc}}(B)$ does not. In
\cite{NaidyukF} the model was proposed that in the mixed state of
type-II superconductor $I_{\textrm{exc}}$ is proportional not only
to $\Delta(B)$, but also to the SC volume, outside vortices. The
size of the contacts in Fig.\,\ref{y9} can be estimated  from
(\ref{wex}) in a few tens of nanometers, while coherence length in
YNi$_{2}$B$_{2}$C is about 5--8\,nm \cite{Muller,Muller1}.
Therefore a number of vortices can penetrate the PC area at a
field approaching $B_{c2}$.

%\section{Discussion}

\section{Conclusion}

We have carried out investigations of the electron-phonon spectral
function and the SC energy gap in YNi$_{2}$B$_{2}$C by PCS. We
have measured PC EPI spectra of YNi$_{2}$B$_{2}$C in the normal
state showing the dominant phonon maximum at about 12\,meV along
with hump or kink around 50\,meV. Position of these features in
the PC spectra corresponds to the maxima in the phonon DOS
measured by neutron diffraction. However most of the phonon peaks
are not resolved in the PC spectra. The reason can be weak
contribution of some phonons into large-angle electron-phonon
scattering in PC and/or large line widths of certain phonon modes.
We did not found appreciable difference in the PC EPI spectra for
PCs demonstrating different SC gap value, what may testify that
the gap variation (anisotropy) is connected with the electronic
structure.

The observed variation of the gap is such that the small gap
($\Delta\sim$1.5\,meV) is characteristic for the direction along
the a-axis, while along the c-axis the gap is higher
($\Delta\sim$2.3\,meV) and similar or slightly larger gap is
measured for the [110] direction. The directional variation
$\Delta$ along with the absence of marked anisotropy of PC EPI
spectra can be an issue of complex multiband ground state in
YNi$_{2}$B$_{2}$C. Multiband scenario for the SC state is also
supported by the observation of the small gap and the excess
current behavior for YNi$_{2}$B$_{2}$C similar as in the
two-gap(band) superconductor MgB$_2$.

\section*{Acknowledgements}

The support of the {\it Deutsche Forschungsgemeinschaft} within
SFB 463 "Rare earth transition metal compounds: structure,
magnetism and transport",  the U.S. Civilian Research and
Development Foundation for the Independent States of the Former
Soviet Union (grant no. UP1-2566-KH-03) and of the National
Academy of Sciences of Ukraine are acknowledged. The
investigations were carried out in part with the help of donated
by A. von Humboldt Foundation (Germany) equipments.


\begin{thebibliography}{10}

\bibitem{Cava}
R. J. Cava, H. Takagi, H. W. Zandbergen, J. J. Krajewski, W. F. Peck
Jr., T. Siegrist, B. Batlogg, R. B. van Dover, R. J. Felder, K.
Mizuhashi, J. O. Lee, H. Eisaki, S. Uchida, Nature {\bf 367} 252
(1994).

\bibitem{Nagarajan}
R. Nagarajan, C. Mazumdar, Z. Hossain, S. K. Dhar, K. V.
Gopalakrisnan, L. C. Gupta, C. Godart, B. D. Padalia, R.
Vijayaraghavan, Phys. Rev. Lett. {\bf 72} (1994) 274.

\bibitem{Muller}
K.-H. M\"{u}ller, V. N. Narozhnyi, Rep. Prog. Phys. \textbf{64},
943 (2001).

\bibitem{Muller1}K.-H. M\"{u}ller, G. Fuchs, S.-L. Drechsler and V. N. Narozhnyi,
in: \textit{Magnetic and Superconducting Properties of Rare Earth
Borocarbides of the Type RNi$_2$B$_2$C}, Handbook of Magnetic
Materials, (Ed. K. H. J. Buschow), Elsevier North-Holland, Vol.
14, (2002), pp. 199-305.

\bibitem{Drechsler}
S.-L. Drechsler, S. V. Shulga, K.-H. M\"{u}ller, G. Fuchs, J.
Freudenberger, G. Behr, H. Eschrig, L. Schultz, M. S. Golden, H.
von Lips, J. Fink, V. N. Narozhnyi, H. Rosner, P. Zahn, A. Gladun,
D. Lipp, A. Kreyssig, M. Loewenhaupt, K. Koepernik, K. Winzer, K.
Krug, Physica C \textbf{317-318}, 117 (1999).

\bibitem{Maki} K. Maki, P. Thalmeier, and H. Won, Phys.\ Rev. B
{\textbf{65}}, 140502 (2002).

\bibitem{Izawa}
K.\ Izawa, K.\ Kamata, Y.\ Nakajima, Y.\ Matsuda, T.\ Watanabe,
M.\ Nohara, H.\ Takagi, P.\ Thalmeier, K.\ Maki, Phys.\ Rev.\ Lett.\
{\textbf{89}}, 137006 (2002).

\bibitem{Park}
T.\ Park, E. E. M.\ Chia, M. B.\ Salamon, E. D.\ Bauer, I.\
Vekhter, J. D.\ Thompson, E. M.\ Choi, H. J. Kim, S.-I. Lee, P. C.
Canfield, Phys. Rev. Lett. \textbf{92}, 237002 (2004).

\bibitem{Martinez} P. Martinez-Samper, H. Suderow, S. Vieira, J. P. Brison, N.
Luchier, P. Lejay, and P. C. Canfield, Phys. Rev. B 67, 014526
(2003).

\bibitem{Pratap}
P. Raychaudhuri, D. Jaiswal-Nagar, Goutam Sheet, S. Ramakrishnan,
and H. Takeya, Phys. Rev. Lett. {\bf 93}, 156802 (2004).
%Evidence of Gap Anisotropy in Superconducting YNi$_{2}$B$_{2}$C
%Using Directional Point-Contact Spectroscopy.

\bibitem{Mukho}S. Mukhopadhyay, Goutam Sheet, P. Raychaudhuri, H. Takeya, Phys.
Rev. B {\bf 72}, 014545 (2005).

\bibitem{Bashlakov}
D. L. Bashlakov, Yu. G. Naidyuk, I. K. Yanson, S. C. Wimbush, B.
Holzapfel, G. Fuchs and S-L Drechsler, Supercond. Sci. Technol.
{\bf 18}, 1094 (2005).

\bibitem{Yokoya} T. Yokoya, T. Baba, S. Tsuda, T. Kiss, A. Chainani, S. Shin,
T. Watanabe, M. Nohara, T. Hanaguri, H. Takagi, Y. Takano, H. Kito,
J. Itoh, H. Harima, T. Oguchi, J. of Physics and Chemistry of Solids
{\bf 67}, 277 (2006).

\bibitem{Huang} C. L. Huang, J.-Y. Lin, C. P. Sun, T. K. Lee, J. D. Kim, E. M.
Choi, S. I. Lee, and H. D. Yang,  Phys. Rev. B \textbf{73}, 012502
(2006).

\bibitem{drechsler04}
S.-L. Drechsler, H.\ Rosner, I.\ Opahle, and H.\ Eschrig, Physica
C \textbf{408}, 104 (2004).

\bibitem{drechsler03}
S.-L.\ Drechsler, I.\ Opahle, S. V.\ Shulga, H.\ Eschrig, G.\
Fuchs, K.-H.\ M\"{u}ller, W.\ L\"{o}ser, H.\ Bitterlich, G.\ Behr,
and H.\ Rosner, Physica B \textbf{329}, 1352 (2003).

\bibitem{drechsler01}
S.-L.\ Drechsler, H.\ Rosner, S. V.\ Shulga, I.\ Opahle, H.\
Eschrig, J.\ Freudenberger, G.\ Fuchs, K.\ Nenkov, K.-H.\
M\"{u}ller, H. Bitterlich, W. L\"{o}ser, G.\ Behr, D.\ Lipp, and A.\
Gladun, Physica C \textbf{364-365} 31 (2001).

\bibitem{Naid}
Yu. G.\ Naidyuk and I.K.\ Yanson, \textit{Point-Contact
Spectroscopy}, Springer Series in Solid-State Sciences, Vol.145
(Springer Science+Business Media, Inc, 2005).

\bibitem{KOS}
I. O. Kulik, A. N. Omelyanchouk, R. I. Shekhter, Fiz. Nizk. Temp.
\textbf{3}, 1543 (1977) {[}Sov. J. Low Temp. Phys. \textbf{3}, 840
(1977){]}.

\bibitem{Kulik}
I. O. Kulik, Fiz. Nizk. Temp. \textbf{18}, 450 (1992) {[}Sov. J.
Low Temp. Phys. \textbf{18}, 302 (1992){]}.

\bibitem{BTK82} G. E. Blonder, M. Tinkham, T. M. Klapwijk, Phys. Rev. B
{\bf25}, 4515 (1982).

\bibitem{Ryba1}
L. F. Rybaltchenko, I. K. Yanson, A. G. M. Jansen, P. Mandal, P.
Wyder, C. V. Tomy, D. McK Paul, Physica B \textbf{218}, 189
(1996).

\bibitem{Yanson0}
I. K. Yanson, V. V. Fisun,  A. G. M. Jansen, P. Wyder, P. C.
Canfield, B. K. Cho, C. V. Tomy  and D. McK Paul, Phys. Rev. Lett.
{\bf 78} 935 (1997).

\bibitem{Yanson1}
I. K. Yanson, V. V. Fisun, A. G. M. Jansen, P. Wyder, P. C.
Canfield, B. K. Cho, C. V. Tomy, D. M. Paul, Fiz. Nizk. Temp.
\textbf{23}, 951 (1997) {[} Low Temp. Phys. \textbf{23}, 712
(1997)].

\bibitem{Gompf}
F.\ Gompf, W.\ Reichardt, H.\ Schober, B.\ Renker, M.\
Buchgeister, Phys. Rev. B \textbf{55}, 9058 (1997).

\bibitem{M2Sconf}  Yu. G.\ Naidyuk, D. L.\ Bashlakov, I.  K.\
Yanson, G.\ Fuchs, G.\ Behr,  D.\ Souptel, and S.-L. Drechsler, to
be published in Physica C (2007).

\bibitem{Souptel}
D. Souptel, G. Behr, W. L\"{o}ser, K. Nenkov, G. Fuchs, J. of Crystal
Growth, \textbf{275}, e91 (2005).

\bibitem{Wexler} A.~Wexler, Proc.~Phys.~Soc. (London) {\bf 89}, 927
(1966).

\bibitem{Shulga}
S. V. Shulga, S.-L. Drechsler, G. Fuchs, K.-H. M\"{u}ller, K.
Winzer, M. Heinecke, and K. Krug, Phys. Rev. Lett. {\bf 80},
1730--1733 (1998).

\bibitem{Kulik1}
I. O. Kulik, Fiz. Nizk. Temp. \textbf{11}, 937 (1985) {[}Sov. J.
Low Temp. Phys. \textbf{11}, 516 (1985){]}.

\bibitem{Shekhter83}
R. I. Shekhter and I. O. Kulik, Fiz. Nizk. Temp. {\bf 9}, 46
(1983) {[}Sov. J. Low Temp. Phys. {\bf 9}, 22 (1983){]}.

\bibitem{Dynes78}
R. C. Dynes, V. Naraynamurti, J. P. Garno, Phys. Rev. Lett. {\bf
41}, 1509 (1978).

%\bibitem{shulga}
%S. V.\ Shulga and S.-L. Drechsler, J.\ Low Temp. Phys. (2001).

\bibitem{NaidSCES}
Yu. G. Naidyuk, O. E. Kvitnitskaya, I. K. Yanson, G. Fuchs, K. Nenkov,
S.-L. Drechsler, G. Behr, D. Souptel, K.\ Gloos, Physica B
\textbf{359-361}, 469 (2005).

\bibitem{Reich}
W. Reichardt,  R. Heid, and K. P. Bohnen, J. of Superconductivity,
\textbf{18}, 759 (2005).

%\bibitem{Rathna}
%K. D. D.~Rathnayaka, D. G.~Naugle, B. K.~Cho, P. C.~Canfield, Phys.
%Rev. B \textbf{53}, 5688 (1996).

\bibitem{Bobrov} N. L. Bobrov, S. I. Beloborod`ko, L. V. Tyutrina, V. N. Chernobay,
I. K. Yanson, D. G. Naugle and K. D. D. Rathnayaka, Fiz. Nizk. Temp.
\textbf{32}, 641 (2006) [Low Temp. Phys. \textbf{32}, 489 (2006)].

\bibitem{Szabo} P. Szab\'o, P. Samuely, J. Ka\v{c}mar\v{c}ik, T. Klein,  J. Marcus,
D. Fruchart, S. Miraglia, C. Marcenat, and A. G. M. Jansen, Phys.
Rev. Lett. {\bf 87}, 137005 (2001).

\bibitem{Baba}T. Baba, T. Yokoya, S. Tsuda, T. Kiss, T. Shimojima, S.
Shina, T. Togashi, C. T. Chen, C. Q. Zhang, S. Watanabe, T.
Watanabea, M. Nohara, H. Takagi, Physica B {\bf 378-380}, 469
(2006).

\bibitem{NaidyukF}
Yu. G. Naidyuk, O. E. Kvitnitskaya, I. K. Yanson, S. Lee, and S.
Tajima, Solid State Commun. \textbf{133}, 363 (2005).

\bibitem{Artemenko}
S. N. Artemenko, A. F. Volkov and A. V. Zaitsev, Solid State
Communs. \textbf{30}, 771 (1979).

\bibitem{Belob}
S. I. Beloborod'ko, A. N. Omelyanchouk, Fiz. Nizk. Temp.
\textbf{17}, 994 (1991) [Sov. J. Low Temp. Phys. \textbf{17}, 518
(1991)].

\end{thebibliography}
\end{document}